\documentclass[12pt]{article}
\usepackage{graphics}
\usepackage{amsthm}
\usepackage{amssymb}
\usepackage{amsmath}
\usepackage{amsfonts}
\usepackage{epic}
\usepackage{hyperref}
\usepackage{epsfig}
\usepackage{bm}
\usepackage{amscd}

\theoremstyle{plain}

\theoremstyle{definition}

\def\be{\begin{equation}}
\def\ee{\end{equation}}
\smallskip
\input{tcilatex}

\begin{document}

\begin{titlepage}
\begin{flushright}
%hep-th/.......\\
\end{flushright}
%%%%%%%%%%%%%%%%%%%%%%%%%%%%%%%%%%%%%%%%%%%%%%%%%%%%%%%%%%%%%%%%%%%%%%%%
\begin{center}
\noindent{{\LARGE{Three-Point Function in Perturbed Liouville Gravity}}}

\smallskip
\smallskip
\noindent{\large{Gaston Giribet}}
\end{center}
\smallskip
\smallskip
\centerline{Department of Physics, Universidad de Buenos Aires}
\centerline{{\it Ciudad Universitaria, 1428. Buenos Aires, Argentina.}}
\smallskip
\centerline{and}
\smallskip
\centerline{Institute of Physics, Universidad Nacional de La Plata}
\centerline{{\it C.C. 67, 1900, La Plata, Argentina.}}
\smallskip

\smallskip

\begin{abstract}
Three-point correlation function in perturbed conformal field
theory coupled to two-dimensional quantum gravity (perturbed
Liouville gravity) is explicitly computed by using the free field
approach. The representation considered here is the one recently
proposed in \cite{Yo}  to describe the string theory in $AdS_3$
space. Consequently, this computation extends previous results
which presented free field calculations of particular cases of string
amplitudes, and confirms that the free field approach
leads to the exact result. Remarkably, this representation allows
to compute winding violating three-point functions without making use of the
spectral flow operator. Exact agreement is obtained with previous computations 
of these observables, which were done by following different formalisms.

\end{abstract}

\end{titlepage}
%%%%%%%%%%%%%%%%%%%%%%%%%%%%%%%%%%%%%%%%%%%%%%%%%%%%%%%%%%%%%%%%%%%%%%%%

%\newpage

%\tableofcontents

\newpage

%%%%%%%%%%%%%%%%%%%%%%%%%%%%%%%%%%%%%%%%%%%%%%%%%%%%%%%%%%%%%%%%%%%%%%

\section{Introduction}

In a recent paper \cite{Yo}, a new free field representation of string
theory in $AdS_{3}$ was introduced in order to realize the explicit
identities that, according to what was proven in \cite{R}, turn out to
connect the correlation functions in both Liouville and $SL(2)_{k}$ WZNW
theories. Such representation corresponds to a perturbed conformal field
theory coupled to two-dimensional quantum gravity (perturbed Liouville
gravity). Then, this enables to make use of all what we have learned about
Liouville field theory and then gain information about the WZNW model. The
purpose of this brief note is that of emphasizing the usefulness of such
realization by explicitly showing how the free field approach can be used to
compute three-point scattering amplitudes in $AdS_{3}$. It is known that the
free field approach and the Coulomb gas-like prescription were previously
employed to this end \cite{BB}; however, the computation here regards those
cases that were not worked out in previous free field calculations. Our
attention will be focussed on the three-point functions that violate the
winding number conservations. In fact, even though free field computations
of such observables were previously presented in the literature \cite{GN3},
it was done by assuming some kind of kinematic restriction, $e.g.$ the
assumption that one of the incoming strings was represented by a
highest-weight state of the $SL(2,R)_{k}$ representations. Moreover,
previous free field computations also considered particular relations
between left-moving $m$ and right-moving $\overline{m}$ momenta, imposing in
such a way certain constraints on the angular momentum of the interacting
strings. Here, we relax such assumptions and calculate the generic ``winding
violating'' three-point amplitude in $AdS_{3}$ within the framework of the
Coulomb gas-like prescription. Besides, we are able to compute correlations
involving states of generic winding number, without introducing intricate
tricks for the definition of states with winding number grater than one and,
remarkably, without resorting to the introduction of the spectral flow
operator.

In the following section we briefly review the free field representation
that will be used. In section 3, we compute the three-point amplitude that
violates the winding conservation. We do this in detail, by emphasizing the
steps through the calculation.

\section{Perturbed Liouville gravity}

\subsection{Free field representation}

Let us begin by briefly reviewing the free field representation we will
employ. The action of the model is that of a matter conformal model $S_{M}$
coupled to the Liouville action $S_{L}$. This takes the form

\begin{equation}
S=\frac{1}{4\pi }\int d^{2}z\left( -\partial \varphi \bar{\partial}\varphi
+QR\varphi +\mu e^{\sqrt{2}b\varphi }\right) +S_{M}  \label{action}
\end{equation}%
where $Q=b+b^{-1}$, and we define the convenient notation $b^{-2}=k-2\in 
\mathbb{R}_{>0}$. We will set the value of the Liouville cosmological
constant as $\mu =1$ by properly rescaling the zero mode of $\varphi (z)$
(see \cite{Yu} for an excellent review on Liouville theory). The specific
model representing the ``matter sector'' corresponds to a $c<1$ conformal
field theory defined by the action%
\begin{equation*}
S_{M}=\frac{1}{4\pi }\int d^{2}z\left( \partial X^{0}\bar{\partial}%
X^{0}-\partial X^{1}\bar{\partial}X^{1}-i\sqrt{k}RX^{1}+\Phi _{aux}\right) ,
\end{equation*}%
where the auxiliary field $\Phi _{aux}(z)$ is a perturbation, represented by
a relevant primary operator of the matter sector and properly dressed with
the coupling to the Liouville field in order to turn it into a marginal
deformation. This takes the form%
\begin{equation}
\Phi _{aux}(z)=(1/c_{k})\quad e^{-\sqrt{\frac{k-2}{2}}\varphi (z)+i\sqrt{%
\frac{k}{2}}X^{1}(z)},
\end{equation}%
where $c_{k}$ is simply a $k$-dependent numerical factor, see \cite{R} for
details. This is a perturbed CFT coupled to Liouville gravity in the spirit
of the models studied in Ref. \cite{Z}. The stress tensor of the theory is
then given by 
\begin{equation*}
T(z)=-\frac{1}{2}\left( \partial \varphi \right) ^{2}+\frac{Q}{\sqrt{2}}%
\partial ^{2}\varphi -\frac{1}{2}\left( \partial X^{1}\right) ^{2}-i\sqrt{%
\frac{k}{2}}\partial ^{2}X^{1}+\frac{1}{2}\left( \partial X^{0}\right) ^{2},
\end{equation*}%
and leads to the central charge%
\begin{equation*}
c=\frac{3k}{k-2}.
\end{equation*}%
The fields $X^{0}(z)$ and $X^{1}(z)$ have time-like and space-like
signatures respectively; namely%
\begin{equation*}
\left\langle X^{0}(z_{1})X^{0}(z_{2})\right\rangle =-\left\langle
X^{1}(z_{1})X^{1}(z_{2})\right\rangle =2\ln |z_{1}-z_{2}|.
\end{equation*}

Auxiliary field $\Phi _{aux}(z)$ enters in the action as an interaction
term, involving the Liouville field $\varphi (z)$ and coupling it with the
field $X^{1}(z)$. From the viewpoint of the computation of correlation
functions, both the operator $\Phi _{aux}(z)$ and the cosmological term $\mu
e^{\sqrt{2}b\varphi (z)}$ play the role of screening charges in the Coulomb
gas-type realization. Actually, these are $(1,1)$-operators of the theory.
Then, different amounts of both operators would be required for the
correlation functions to be non-vanishing. However, we will focus the
attention to those correlators that do not involve insertion of the
perturbation field $\Phi _{aux}(z)$. These cases lead to the violation of
winding number conservation. A similar free field realization was
independently considered in \cite{F}.

The vertex operators in the theory are given by 
\begin{equation}
\Phi _{j,m,\bar{m}}^{\omega }(z)=\frac{c_{k}\Gamma (-m-j)}{\Gamma (j+1+\bar{m%
})}e^{\sqrt{\frac{2}{k-2}}(j+\frac{k}{2})\varphi (z)+i\sqrt{\frac{2}{k}}(m-%
\frac{k}{2})X^{1}(z)-i\sqrt{\frac{2}{k}}(m+\frac{k}{2}\omega
)X^{0}(z)}\times h.c.
\end{equation}%
where $h.c.$ stands for the anti-holomorphic part, which also contains the
dependence on $\bar{m}$. It is worth pointing out that the normalization $%
\frac{c_{k}\Gamma (-m-j)}{\Gamma (j+1+\bar{m})}$ is the precisely the one
required in order to reproduce the one-to-one correspondence between
correlation functions in WZNW theory and Liouville theory. These are primary
operators and have conformal dimension%
\begin{equation*}
h_{j,m,\omega }=-\frac{j(j+1)}{k-2}-m\omega -\frac{k}{2}\omega ^{2}.
\end{equation*}%
This yields the mass spectrum of the theory through the Virasoro constraint $%
h_{j,m,\omega }=1$. On the other hand, the energy of the string states is
given by the quantity $E=m+\overline{m}+k\omega $, which includes both
kinetic and winding contributions. Now, we move to the correlation functions
involving these states.

\subsection{Particular correlation functions}

Here, we are interested in particular $N$-point correlation functions in the
theory. These are denoted as%
\begin{equation*}
A_{j_{1},j_{2},...j_{N};m_{1},m_{2},...m_{N}}^{\omega _{1},\omega
_{2},...\omega _{N}}=\left\langle \Phi _{j_{1},m_{1},\bar{m}_{1}}^{\omega
_{1}}(z_{1})\Phi _{j_{2},m_{2},\bar{m}_{2}}^{\omega _{2}}(z_{2})...\Phi
_{j_{N},m_{N},\bar{m}_{N}}^{\omega _{N}}(z_{N})\right\rangle
\end{equation*}%
and are those satisfying the particular relation $\omega _{1}+\omega
_{2}+...+\omega _{N}=2-N$. According to the free field realization described
in \cite{Yo}, these observables admit an integral representation of the form 
\begin{equation*}
A_{j_{1},j_{2},...j_{N};m_{1},m_{2},...m_{N}}^{\omega _{1},\omega
_{2},...\omega _{N}}=\frac{\Gamma (-s)}{\left( c_{k}\right) ^{2-N}}%
\prod\limits_{d=1}^{N}\frac{\Gamma (-m_{d}-j_{d})}{\Gamma (j_{d}+1+\bar{m}%
_{d})}\prod\limits_{r=1}^{s}\int d^{2}w_{r}\left(
\prod\limits_{n=1}^{s}\prod\limits_{a=1}^{N}|z_{a}-w_{n}|^{\frac{4}{2-k}%
(j_{a}+\frac{k}{2})}\right. \times
\end{equation*}%
\begin{equation}
\times \left. \prod\limits_{l=1}^{s}\prod\limits_{t=1}^{l-1}|w_{l}-w_{t}|^{%
\frac{4}{2-k}}\prod\limits_{b=1}^{N}\prod\limits_{c=1}^{b-1}|z_{b}-z_{c}|^{%
\frac{4}{2-k}(j_{b}+\frac{k}{2})(j_{c}+\frac{k}{2})-2(m_{b}+m_{c}+\omega
_{b}m_{c}+\omega _{c}m_{b})+k(1-\omega _{b}\omega _{c})}\right)  \label{I2}
\end{equation}%
and, as it was mentioned above, correspond to those correlators which do not
receive perturbations of the form $\int d^{2}w\Phi _{aux}(w)$ but merely
contributions of the Liouville screening charge $\int d^{2}we^{\sqrt{2}%
b\varphi (w)}$. These represent ``maximally violating winding'' scattering
amplitudes in $AdS_{3}$ spacetime. In fact, except for the case of the
2-point function, the total winding number is not conserved in such
correlation functions as can be verified from the following conservations
laws 
\begin{eqnarray}
&&\sum_{i=1}^{N}j_{i}+(N-2)\frac{k}{2}+s+1=0  \label{t13} \\
&&\sum_{i=1}^{N}\omega _{i}+N-2=0.  \label{conservation}
\end{eqnarray}%
These conservation laws are due to $\delta (x)$-functions arising in the
integration over the zero modes of the fields $\varphi (z),$ $X^{0}(z)$ and $%
X^{1}(z)$. Besides, correlators that also include ``screenings'' of the type 
$\int d^{2}w\Phi _{aux}(w)$ do satisfy a different compensation relation, in
particular: $\omega _{1}+\omega _{2}+...+\omega _{N}=2-N+M$, where $M$ is
the amount of screening fields $\int d^{2}w\Phi _{aux}(w)$ involved in the
correlators (see \cite{Yo} for details). Then, in the case of the
three-point function, the only non-trivial result including the perturbation
field $\Phi _{aux}(w)$ would be the conserving winding three-point function
which is certainly well known. Let us focus on the non-conservative
amplitude.

\section{The three-point function}

\subsection{Integral representation}

The intention is to compute the three-point function that describes string
scattering amplitudes in $AdS_{3}$ for the case where the conservation of
the total winding number is violated; and we want to do this by using free
fields and without imposing any kinematic restriction on the involved
states. We denote such correlation function as%
\begin{equation*}
A_{j_{1},j_{2},j_{3};m_{1},m_{2},m_{3}}^{\omega _{1},\omega _{2},\omega
_{3}}=\left\langle \Phi _{j_{1},m_{1},\bar{m}_{1}}^{\omega _{1}}(z_{1})\Phi
_{j_{2},\frac{k}{2}-m_{1}-m_{3},\frac{k}{2}-\bar{m}_{1}-\bar{m}%
_{3}}^{-1-\omega _{1}-\omega _{3}}(z_{2})\Phi _{j_{3},m_{3},\bar{m}%
_{3}}^{\omega _{3}}(z_{3})\right\rangle ,
\end{equation*}%
where the quantum numbers are such that satisfy the conservation laws
leading to the non-vanishing result. Then, we will compute it by using the
approach described in \cite{Yo}. By means of the standard techniques of the
Coulomb gas-like prescription, this leads to the following multiple integral
in the whole complex plane%
\begin{equation*}
A_{j_{1},j_{2},j_{3};m_{1},m_{2},m_{3}}^{\omega _{1},\omega _{2},\omega
_{3}}=\Gamma (-s)c_{k}\prod\limits_{a\neq
b}^{3}|z_{a}-z_{b}|^{2(h_{1}+h_{2}+h_{3}-2h_{a}-2h_{b})}\times
\end{equation*}%
\begin{eqnarray}
&&\times \prod\limits_{c=1}^{3}\frac{\Gamma (-m_{c}-j_{c})}{\Gamma (j_{c}+1+%
\bar{m}_{c})}\prod\limits_{r=1}^{s}\int d^{2}w_{r}\left(
\prod\limits_{n=1}^{s}|w_{n}|^{\frac{4}{2-k}(j_{1}+\frac{k}{2})}|1-w_{n}|^{%
\frac{4}{2-k}(j_{2}+\frac{k}{2})}\prod\limits_{l=1}^{s}\prod%
\limits_{t=1}^{l-1}|w_{l}-w_{t}|^{\frac{4}{2-k}}\right) \times  \notag \\
&&\times \delta (m_{1}+m_{2}+m_{3}-k/2)\delta (\bar{m}_{1}+\bar{m}_{2}+\bar{m%
}_{3}-k/2)\delta (s+j_{1}+j_{2}+j_{3}+1+k/2)  \label{I}
\end{eqnarray}%
where $\int d^{2}w_{r}=\frac{1}{2\pi i}\int dw_{r}\int d\overline{w}_{r}$.
The integration over the zero-mode of the fields $\varphi (z)$, $X^{0}(z)$
and $X^{1}(z)$ states that the amount of integrals to be performed is given
by $s=-j_{1}-j_{2}-j_{3}-\frac{k}{2}-1$, while the momenta obey the
conservation laws $m_{1}+m_{2}+m_{3}=\bar{m}_{1}+\bar{m}_{2}+\bar{m}_{3}=%
\frac{k}{2}$. Consequently, the conservation of the winding number is
violated in one unit, namely $\omega _{1}+\omega _{2}+\omega _{3}=-1$.
Notice that the integral (\ref{I}) is a Dotsenko-Fateev integral (similar to
those arising in the minimal models) and can be explicitly solved by using
the results of Ref. \cite{DF}. It is worth pointing out that, as it is usual
within similar contexts, the integral formula of the type (\ref{I2}) has to
be understood formally, and a kind of analytic extension of it is required
in order to construct generic correlators with non-integer $s$. The features
related to such analytic extension are basically two: First, it is evident
that the products of the form $\prod_{n=1}^{s}$ in (\ref{I}) only make sense
for positive integers $s$. Then, the analytic continuation of the formulas
containing such products (after integration) is needed in order to consider
generic values of the momenta (see Ref. \cite{GL} for more details). For
instance, this is similar to what occurs in the computation of correlation
functions in 2D minimal gravity, \cite{D}. The second issue is the presence
of the overall factor $\Gamma (-s),$ which arises after integrating over the
zero mode of the Liouville field $\varphi (z)$. This factor diverges for
positive integers $s,$ and such a divergence is associated to the
non-compactness of the theory, \cite{dFK}. Here, we follow standard paths in
this kind of computation and proceed by assuming an analytic continuation of
the formulas obtained after the integration. Then, we can integrate out (\ref%
{I}) by using the following identity (see Ref. \cite{GN3})

\begin{equation*}
I_{s}(J_{1},J_{2};k)=\prod\limits_{r=1}^{s}\int d^{2}w_{r}\left(
\prod\limits_{n=1}^{n}|w_{n}|^{\frac{4}{k-2}J_{1}-2}|1-w_{n}|^{\frac{4}{k-2}%
J_{2}}\prod\limits_{l=1}^{s}\prod\limits_{t=1}^{l-1}|w_{l}-w_{t}|^{\frac{4}{%
2-k}}\right) =
\end{equation*}%
\begin{eqnarray}
&=&\frac{(k-2)}{\Gamma (-s)}\left( \frac{\pi \Gamma \left( \frac{1}{k-2}%
\right) }{\Gamma \left( 1-\frac{1}{k-2}\right) }\right) ^{s}\frac{\Gamma
(-1-J_{1}-J_{2}-J_{3})\Gamma (2J_{2}+1)\Gamma (J_{1}-J_{2}-J_{3})\Gamma
(-J_{1}-J_{2}+J_{3})}{\Gamma (2+J_{1}+J_{2}+J_{3})\Gamma (-2J_{2})\Gamma
(1-J_{1}+J_{2}+J_{3})\Gamma (1+J_{1}+J_{2}-J_{3})}\times  \notag \\
&&\times \frac{%
G_{k}(-2-J_{1}-J_{2}-J_{3})G_{k}(-1-J_{1}+J_{2}-J_{3})G_{k}(-1+J_{1}-J_{2}-J_{3})G_{k}(-1-J_{1}-J_{2}+J_{3})%
}{G_{k}(-1)G_{k}(-2J_{1}-1)G_{k}(-2J_{1}-1)G_{k}(-2J_{3}-1)}\times  \notag \\
&&  \label{I3}
\end{eqnarray}%
where $J_{3}$ has been defined by $J_{3}=s-J_{1}-J_{2}-1$, and where the
special function $G_{k}(x)$ is defined through%
\begin{equation*}
G_{k}(x)=(k-2)^{\frac{x(k-1-x)}{2(k-2)}}\Gamma _{2}(-x|1,k-2)\Gamma
_{2}(k-1+x|1,k-2),
\end{equation*}%
where the Barnes function $\Gamma _{2}(x|1,y)$ is given by%
\begin{equation*}
\ln \Gamma _{2}(x|1,y)=\lim_{\varepsilon \rightarrow 0}\frac{d}{d\varepsilon 
}\sum\limits_{n=0}^{\infty }\sum\limits_{m=0}^{\infty }\left(
(x+n+my)^{-\varepsilon }-(1-\delta _{n,0}\delta _{m,0})(n+my)^{-\varepsilon
}\right)
\end{equation*}%
where the presence of the factor $(1-\delta _{n,0}\delta _{m,0})$ in the
right hand side means that the sum in the second term does not take into
account the step $m=n=0$.

Some useful functional relations of these functions are the following%
\begin{eqnarray}
G_{k}(x) &=&G_{k}(-x-k+1)  \label{uno} \\
G_{k}(x) &=&G_{k}(x+1)\gamma \left( 1+\frac{1+x}{k-2}\right)  \label{dos} \\
G_{k}(x) &=&G_{k}(x-k+2)(k-2)^{2x+1}\gamma (-x)  \label{tres}
\end{eqnarray}%
where we have made use of the standard notation%
\begin{equation*}
\gamma (x)=\frac{\Gamma (x)}{\Gamma (1-x)}.
\end{equation*}%
The $G_{k}(x)$ function develops simple poles at $x=p+q(k-2)$ and $%
x=-1-p-(1+q)(k-2)$, for $p,q\in {\mathbb{Z}}_{\geq 0}$. The functional
properties (\ref{uno})-(\ref{tres}), due to the fact that these involve the $%
\gamma (x)$ function as well, can be used to prove the above integral
formula for $I_{s}(J_{1},J_{2};k)$ in the case one prefers starting with the
Dotsenko-Fateev expression in terms of product of $\Gamma (x)$ functions
(see Appendix of Ref. \cite{DF}). The important point here is that the
integral $I_{s}(J_{1},J_{2};k)$ precisely agrees with the one we have to
compute through the identification $J_{1}=-1-j_{1}$, $J_{2}=-\frac{k}{2}%
-j_{2}$ and $J_{3}=-1-j_{3}$. Then, we are ready to evaluate the three-point
function. First, notice that the relations (\ref{uno})-(\ref{tres}) help us
in writing%
\begin{eqnarray*}
G_{k}(-1-j_{1}+j_{2}+j_{3}+k/2) &=&\frac{G_{k}(j_{1}-j_{2}-j_{3}-k/2)}{%
\gamma (-j_{1}+j_{2}+j_{3}+k/2)}(k-2)^{k-1-2(j_{1}-j_{2}-j_{3})}, \\
G_{k}(-1+j_{1}+j_{2}-j_{3}+k/2) &=&\frac{G_{k}(-j_{1}-j_{2}+j_{3}-k/2)}{%
\gamma (j_{1}+j_{2}-j_{3}+k/2)}(k-2)^{k-1+2(j_{1}+j_{2}-j_{3})}
\end{eqnarray*}%
and also%
\begin{equation*}
G_{k}(2j_{2}+k-1)=G_{k}(1-2j_{2}-k)(k-2)^{2(2j_{2}+k-1)}\gamma
(1-2j_{2}-k)\gamma \left( 1-\frac{2j_{2}+k-1}{k-2}\right) .
\end{equation*}%
Since we are mainly interested in the string theory applications (and
consequently in ``correlation numbers'' instead of correlation functions) we
can make use of the projective invariance and set the worldsheet inserting
points as usual: $z_{1}=0,$ $z_{2}=1$ and $z_{3}=\infty $. By integrating
out and using the functional relations (\ref{uno})-(\ref{tres}) we find the
following expression for the violating three-point correlation number%
\begin{equation*}
A_{j_{1},j_{2},j_{3};m_{1},m_{2},m_{3}}^{\omega _{1},\omega _{2},\omega
_{3}}=(k-2)\left( \pi \gamma \left( \frac{1}{k-2}\right) \right)
^{-j_{1}-j_{2}-j_{3}-\frac{k}{2}-1}\frac{c_{k}\Gamma (-m_{1}-j_{1})\Gamma
(-m_{2}-j_{2})\Gamma (-m_{3}-j_{3})}{\Gamma (j_{1}+1+\bar{m}_{1})\Gamma
(j_{2}+1+\bar{m}_{2})\Gamma (j_{3}+1+\bar{m}_{3})}\times
\end{equation*}%
\begin{eqnarray}
&&\times \frac{G_{k}(j_{1}+j_{2}+j_{3}+\frac{k}{2})G_{k}(-j_{1}-j_{2}+j_{3}-%
\frac{k}{2})G_{k}(j_{1}-j_{2}-j_{3}-\frac{k}{2})G_{k}(1+j_{1}-j_{2}+j_{3}-%
\frac{k}{2})}{\gamma \left( -j_{1}-j_{2}-j_{3}-\frac{k}{2}\right) \gamma
\left( -\frac{2j_{2}+1}{k-2}\right)
G_{k}(-1)G_{k}(2j_{1}+1)G_{k}(1-k-2j_{2})G_{k}(2j_{3}+1)}\times  \notag \\
&&\times \delta (m_{1}+m_{2}+m_{3}-k/2)\delta (\bar{m}_{1}+\bar{m}_{2}+\bar{m%
}_{3}-k/2)\delta (s+j_{1}+j_{2}+j_{3}+1+k/2).  \label{result}
\end{eqnarray}%
where we preferred writing this in such a way because it permits to compare
with the results in the literature (although the replacement $%
j_{i}\rightarrow -j_{i}$ is still necessary to compare with Ref. \cite{MO3}%
). In fact, this formula exactly agrees with the one found in the literature
by using rather different approaches (see also \cite{Yo2}). Besides, this
extends previous computations which were done by using free field techniques
because it does represent the ``generic'' three-point violating winding
amplitude in $AdS_{3}$. Notice that this computation does not require the
insertion of the spectral flow operator (conjugate representations of the
identity operator) and seems to be valid for states with generic winding
number (spectral flow parameter $\omega $). In particular, the fact that
this computation did not make use of the spectral flow operator is actually
interesting. The inclusion of such additional vertex in the correlators with
the purpose of realizing the violation of the winding number is actually one
of the most ingenious tricks; however, from the viewpoint of the standard
prescription for computing correlation functions, the introduction of such
an operator could appear as a little heterodox; then, having found an
alternative way of calculating seems to be a good point. Besides, it is
worth mentioning that the formula above is consistent with the FZZ
conjecture (cf. Ref. \cite{FH}).

\subsection{Remarks on the pole structure}

Some remarks are in order: First, besides the usefulness of expression (\ref%
{result}) in order to compare with the results of \cite{MO3} and \cite{Yo2},
the result can be also written in a way such that the symmetry under
interchanges $j_{i}\leftrightarrow j_{j}$ for $i,j\in \{1,2,3\}$ turns out
to be explicit. By using the relations (\ref{uno}) and (\ref{dos}) we can
write (\ref{result}) in the following form, where such symmetry manifestly
appears,%
\begin{equation*}
A_{j_{1},j_{2},j_{3};m_{1}m_{2},m_{3}}^{\omega _{1},\omega _{2},\omega
_{3}}=(k-2)\left( \pi \gamma \left( \frac{1}{k-2}\right) \right)
^{-\sum_{a=1}^{3}j_{a}-\frac{k}{2}-1}\prod_{b=1}^{3}\frac{\Gamma
(-m_{b}-j_{b})G_{k}(2j_{b}-\sum_{a=1}^{3}j_{a}-k/2)}{\Gamma (j_{b}+1+\bar{m}%
_{b})G_{k}(2j_{b}+1)}\times
\end{equation*}%
\begin{equation*}
\times \frac{c_{k}\gamma
(1+\sum_{a=1}^{3}j_{a}+k/2)G_{k}(\sum_{a=1}^{3}j_{a}+k/2)}{G_{k}(-1)}\delta
(\sum_{a=1}^{3}m_{a}-k/2)\delta (\sum_{a=1}^{3}\bar{m}_{a}-k/2)\delta
(s+\sum_{a=1}^{3}j_{a}+1+k/2).
\end{equation*}%
On the other hand, notice that we can obtain the two-point function by
properly performing the limit $j_{2}\rightarrow -k/2$ in the expression for
the three-point function we just obtained. This is because the 2-point
function does conserve the winding number. In fact, by taking into account
the functional relation%
\begin{equation*}
\lim_{\varepsilon \rightarrow 0}\frac{G_{k}(\varepsilon -x)G_{k}(\varepsilon
+x)}{G_{k}(2\varepsilon +1)}=-2\pi i(k-2)G_{k}(-1)\gamma \left( 1+\frac{1}{%
k-2}\right) \delta (x)
\end{equation*}%
and using (\ref{dos}) we find that in the limit $\varepsilon
=-j_{2}-k/2\rightarrow 0$ the expression (\ref{tres}) reduces to%
\begin{equation*}
A_{j_{1},,j_{3};m_{1},m_{3}}^{\omega _{1},\omega _{3}}=-2\pi
i(k-2)^{2}\left( \pi \gamma \left( \frac{1}{k-2}\right) \right) ^{-2j_{1}-1}%
\frac{c_{k}\gamma \left( 2j_{1}+1\right) \Gamma (-m_{1}-j_{1})\Gamma
(m_{1}-j_{1})}{\gamma \left( -\frac{2j_{1}+1}{k-2}\right) \Gamma (j_{1}+1+%
\bar{m}_{1})\Gamma (j_{1}+1-\bar{m}_{1})}\times
\end{equation*}%
\begin{equation}
\times \delta (m_{1}+m_{3}-k/2)\delta (\bar{m}_{1}+\bar{m}_{3}-k/2)\delta
(j_{1}-j_{3}).  \label{2p}
\end{equation}%
This is, up to a $k$-dependent factor, the reflection coefficient, and is
non vanishing only for the cases fulfilling the conditions $m_{1}+m_{3}=\bar{%
m}_{1}+\bar{m}_{3}=\omega _{1}+\omega _{3}=0$.

Other comment regards the operator product expansion. The OPE and,
consequently, the fusion rules of the theory are codified in the pole
structure of the three-point function. The OPE for the $\omega =0$ sector of
the Hilbert space was studied in detail in Ref. \cite{S} and was analyzed in
relation with the four-point function in Ref. \cite{MO3}. Here, we want to
make a few remarks on the mixing between sectors $\omega =0$ and $\omega =1$%
. Let us consider the short distance behavior 
\begin{eqnarray}
\Phi _{j_{1},m_{1}}^{\omega _{1}=0}(z_{1})\Phi _{j_{2},m_{2}}^{\omega
_{2}=0}(z_{1}) &\simeq &\sum_{\omega }\int_{C}dj\ dm\ d\overline{m}\quad
|z_{1}-z_{2}|^{2(h_{j,m,\omega }-h_{j_{1},m_{1},\omega
_{1}}-h_{j_{2},m_{2},\omega _{2}})}\times  \notag \\
&&\times Q_{k}(j_{1},j_{2},j;m_{1},m_{2},m;\omega )\{\Phi _{j,m}^{\omega
}(z_{1})\}+...  \label{OPE}
\end{eqnarray}%
where the dots \textquotedblleft ...\textquotedblright\ stand for
\textquotedblleft other contributions\textquotedblright , and where the
coefficient $Q_{k}(j_{1},j_{2},j;m_{1},m_{2},m;\omega )$ is given by a
quotient between the structure constant (\ref{result}) and the reflection
coefficient (\ref{2p}) of two states with winding number $\omega =1$. To be
precise, a change of sign in such expression also appears because of a
replacing $m_{3}\rightarrow -m$. Because the pole structure of the structure
constants determine the OPE, the arising of the factor $\gamma
^{-1}(-j_{1}-j_{2}-j_{3}-k/2)$ in (\ref{result}) turns out to be important
since it cancels a simple pole coming from the function $%
G_{k}(j_{1}+j_{2}+j_{3}+k/2)$. The sum $\sum_{\omega }$ over the quantum
number ${\omega }$ stands for making explicit that the fusion rules can lead
to the mixing of sectors due to the spectral flow symmetry and eventually
yield the violation of the winding conservation up to one in the three-point
function; accordingly, $\omega \in \{0,\pm 1\}$. On the other hand, the
region of integration $C$, schematically represented in the formal sum $%
\int_{C}dj$ $dm$ $d\overline{m}$, is defined in such a way that the
integration over the indices $j\in -\frac{1}{2}+i\mathbb{R}$ and $\alpha \in
\lbrack 0,1)$ of the continuous series $\mathcal{C}_{j}^{\alpha ,\omega }$
is performed, and so for the contributions due to the poles corresponding to
states of the discrete series $\mathcal{D}_{j}^{\pm ,\omega =1}$. I.e. the
definition of $C$ is understood as running over the sets $\mathcal{C}%
_{j}^{\alpha ,\omega }=\{j,m$ / $j\in -\frac{1}{2}+i\mathbb{R}$, $\alpha \in
\lbrack 0,1)$, $m\in \alpha +{\mathbb{Z}}_{\geq 0}\}$ and encloses the poles
belonging to the sets $\mathcal{D}_{j}^{\pm ,\omega }=\{j,m$ / $j\in \mathbb{%
R}_{<-\frac{1}{2}}$, $m=\pm (j-n)$, $n\in {\mathbb{Z}}_{\geq 0}\}$. These
sets parameterize the (universal covering of the) unitary representations of 
$SL(2,\mathbb{R})$ that are relevant for the string theory applications. For
\textquotedblleft picking up\textquotedblright\ the poles corresponding to
the discrete states contributions, the contours included in $C$ have to be
properly chosen and a regularization procedure is required in those cases
where different poles turn out to coincide, \cite{S}. Besides, the sum over
the quantum numbers $j$, $m$, $\overline{m}$ and $\omega $ in the OPE (\ref%
{OPE}) has to take into account the fact that certain states of discrete
representations of both sectors $\omega =0$ and $\omega =1$ are related one
each other through the identification $\mathcal{D}_{j}^{\pm ,\omega =1}\sim 
\mathcal{D}_{-k/2-j}^{\mp ,\omega =0}$, similarly as what occurs in the
compact $SU(2)_{k}$ case. Besides, a lower bound on the sum over $j$ is
required in order to guarantee the unitarity of the spectrum; namely $2j>1-k$%
. In the case on which we were interested here, unlike the case when the OPE
is considered as being closed among the states of sector $\omega =0$, it is
not necessary to distinguish between discrete $\mathcal{D}_{j}^{\pm ,\omega
} $ and continuous series $\mathcal{C}_{j}^{\alpha ,\omega }$ in order to
analyze the $m$-dependent pole structure of $%
Q_{k}(j_{1},j_{2},j;m_{1},m_{2},m;1)$. This is due to the fact that,
remarkably, the dependence of the violating winding amplitude (\ref{result})
on the parameters $m$ and $\overline{m}$ turns out to be substantially
simpler than the one that corresponds to the winding conserving case. This
is explained by the fact that the field $\Phi _{aux}(z)$ depends on $%
X^{1}(z) $ as well. Hence, the whole pole structure of $%
Q_{k}(j_{1},j_{2},j;m_{1},m_{2},m;1)$ is basically given by the poles of (%
\ref{result}) and by the poles of the $\Gamma (x)$-functions (occurring at $%
x\in {\mathbb{Z}}_{<0}$) arising in the denominator of (\ref{2p}). Within
this framework, it would be certainly interesting to extend the study made
in \cite{S} and \cite{HS} for the case of violating amplitudes. This could
help in understanding the factorization properties of the four-point
function in the $SL(2,R)_{k}$ WZNW model. As mentioned before, the OPE was
studied in connection to the four-point function in Ref. \cite{MO3}, where
it was proven that two incoming states belongings to the sector $\omega =0$
can produce intermediate states with both $\omega =0$ and $\omega =1$.
However, further study is necessary to fully understand the factorization of
the four-point function and our hope is that the free field representation
can help in doing this.

\subsection{Remarks on the $sl(2)_{k}$ invariance}

Now, let us make some remarks about the $sl(2)_{k}$ symmetry of the action (%
\ref{action}). Such symmetry should to be present in the theory since what
one is actually doing is asserting the identity between the free field
realization $Liouville$ $\times U(1)\times U(1)$ and the $SL(2,R)_{k}$ WZNW
model. In fact, \emph{ab initio}, we know that this construction actually
presents such $sl(2)_{k}$ symmetry since it turns out to reproduce those
solutions of the Knizhnik-Zamolodchikov equations that Ribault has found in
Ref. \cite{R}. However, even though the solutions we obtain have the
appropriate symmetry, the question arises as to why does it happen if the
Liouville interaction term $e^{\sqrt{2}b\varphi (z)}$ does not seem to have
regular OPE with the $SL(2,R)_{k}$ currents though. To be precise, even
though one knows that the free field representation presented in Ref. \cite%
{Yo}\ turns out to transform properly by construction (it reproduces
solutions of the KZ\ equation), \ it is also true that it is not obvious
that the Liouville interaction term regarded as a screening charge commutes
with the free field representation of the $sl(2)_{k}$ current algebra as one
could naively expect. Again, why does it happen? The answer to this question
yields from noticing that also the vertex operators $\Phi _{j,m,\overline{m}%
}^{\omega }(z)$ do not satisfy the usual OPE that the vectors of the $%
SL(2,R) $ representations satisfy according to the usual picture. In
particular, it is worth noticing that the $m$-dependent overall factor of
such vertex operators plays a crucial role for this condition to hold.
Again, to be precise, let me make the following observation: The
stress-tensor of the free field theory presented here can be thought of as
the Sugawara construction starting from the following generators of the $%
sl(2)_{k}$ affine algebra

\begin{eqnarray*}
J^{\pm }(z) &=&-i\sqrt{\frac{k}{2}}\partial Y^{1}(z)e^{\mp i\sqrt{\frac{2}{k}%
}(Y^{0}(z)+Y^{1}(z))}\pm \sqrt{\frac{k-2}{2}}\partial \rho (z)e^{\mp i\sqrt{%
\frac{2}{k}}(Y^{0}(z)+Y^{1}(z))} \\
J^{3}(z) &=&i\sqrt{\frac{k}{2}}\partial Y^{0}(z)
\end{eqnarray*}%
\bigskip which follow from the free field redefinition \cite{Yo}%
\begin{eqnarray*}
\rho (z) &=&(1-k)\varphi (z)+i\sqrt{k(k-2)}X^{1}(z) \\
Y^{1}(z) &=&(k-1)X^{1}(z)+i\sqrt{k(k-2)}\varphi (z) \\
Y^{0}(z) &=&-X^{0}(z)
\end{eqnarray*}%
Then, as it can be verified, these currents do not have regular OPE with the
Liouville cosmological constant term as one could naively expect. However,
the non trivial point is that this is precisely what makes the $SL(2,R)_{k}$
to be recovered. Namely, these currents do not presents regular OPE with the
Liouville cosmological term $e^{\sqrt{2}b\varphi (z)}$, but these do not
satisfy the usual OPE with the vertex operators $\Phi _{j,m,\overline{m}%
}^{\omega }(z)$ either; and both facts seem to combine in such a way that
explain why the formulas obtained for the correlators by using this free
field representation turn out to be $SL(2,R)_{k}$ invariant. Let me
emphasize that the proof of such $SL(2,R)_{k}$ invariance of the correlators
simply follows from the fact that these exactly solve the KZ\ equation,
since lead to the solutions of \cite{R} with the appropriate normalization
factor, as by means of the Coulomb gas-like prescription in \cite{Yo}.
Furthermore, let us notice that this is precisely one of the two aspects
that make of this free field construction in terms of the product $%
Liouville\times U(1)\times U(1)$ a non trivial one. Namely, the first non
trivial point is the fact that this construction does not seem to follow
from a simple field redefinitions (\emph{i.e.} there is no clear way for
obtaining this tachyonic interaction term through bosonization, for
instance), and the second non trivial point is precisely the use of this non
standard representations $\Phi _{j,m,\overline{m}}^{\omega }$ which, once
combined with the Liouville cosmological term, restores the $SL(2,R)_{k}$
invariance that the correlators one computes manifest.

\section{Conclusion}

By using the free field representation introduced in \cite{Yo}, we have
computed the three-point winding violating amplitude in $AdS_{3}$ for the
generic case, \textit{i.e.} without imposing the highest-weight state
condition $m_{a}\pm j_{a}=0$ on any vertex and without making assumptions on
the angular momenta $m_{a}-\bar{m}_{a}$. Besides, this computation seems to
involve vertex operators of generic winding number $\omega _{a}$, without
resorting to subtle tricks for defining the vertex of sectors $\omega >1$.
Then, it shows that the free field method turns out to be powerful enough to
reproduce the three-point winding violating amplitude on the sphere in
complete agreement with other calculations. Notice that even the factor $%
\gamma ^{-1}\left( -j_{1}-j_{2}-j_{3}-\frac{k}{2}\right) $ has been
reproduced here and the correct $m$-dependent factor has been also obtained.
This result represents a consistency check for the realization proposed in 
\cite{Yo}, which now has shown to be useful to compute string scattering
amplitudes. We emphasize that our result is based on the free field
representation of Ref. \cite{Yo}, which was defined to exactly realize the
solutions of the Knizhnik-Zamolodchikov equation given in Ref. \cite{R}. 
\begin{equation*}
\end{equation*}

\textbf{Acknowledgement:} It is a pleasure to thank the Centro de Estudios
Cient\'{\i}ficos en Valdivia (CECS) for the hospitality during my stay,
where the first part of this work was done; and the Universit\'{e} Libre de
Bruxelles (ULB), where the revised version of the manuscript was written. I
also thank Yu Nakayama for discussions and important remarks. This work was
supported by Universidad de Buenos Aires and CONICET.


\begin{thebibliography}{99}
\bibitem{Yo} G. Giribet, Nucl.Phys. \textbf{B737} (2006) pp. 209-235.

\bibitem{R} S. Ribault, JHEP \textbf{0509} (2005) pp. 045.

\bibitem{BB} K. Becker and M. Becker, Nucl.Phys. \textbf{B418} (1994) pp.
206- 230.

\bibitem{GN3} G. Giribet and C. N\'u\~nez, JHEP \textbf{0106} (2001) pp. 010.

\bibitem{Yu} Yu Nakayama, Int.J.Mod.Phys. \textbf{A19} (2004) pp. 2771-2930.

\bibitem{Z} Al. Zamolodchikov, \textit{Perturbed Conformal Field Theory on
Fluctuating Sphere}, arXiv: hep-th/0508044.

\bibitem{F} V. Fateev, \textit{Relation between Sine-Liouville and Liouville
correlation functions}, unpublished.

\bibitem{DF} V. Dotsenko and V. Fateev, Nucl.Phys. \textbf{B251} (1985) pp.
691.

\bibitem{D} V. Dotsenko, Nucl.Phys. \textbf{B338} (1990) pp. 747; Nucl.Phys. 
\textbf{B358} (1991) pp. 541.

\bibitem{GL} M. Goulian and M. Li, Phys.Rev.Lett. \textbf{66} (1991) pp.
2051-2055.

\bibitem{dFK} P. Di Francesco and D. Kutasov, Nucl.Phys. \textbf{B375}
(1992) pp. 119-172.

\bibitem{MO3} J. Maldacena and H. Ooguri, Phys.Rev. \textbf{D65} (2002) pp.
106006.

\bibitem{Yo2} G. Giribet, Phys.Lett. \textbf{B628} (2005) pp. 148-156.

\bibitem{FH} T. Fukuda and K. Hosomichi, JHEP \textbf{0109} (2001) pp. 003.

\bibitem{S} Y. Satoh, Nucl.Phys. \textbf{B629} (2002) pp. 188-208.

\bibitem{HS} K. Hosomichi and Y. Satoh, Mod.Phys.Lett. \textbf{A17} (2002)
pp. 683-693.
\end{thebibliography}
\end{document}